\documentclass[12pt, a4paper]{amsart}

\usepackage{amsmath}
\usepackage{amssymb}
\usepackage{amsthm}
\usepackage{graphicx}
\usepackage{color}

\title{DEMON{\tiny{ic}} Dominoes\\ \tiny{Measuring the Speed of the Domino Effect from Sound Recordings}}

\author{Ron Larham*}
\thanks{*BAE Systems, Broad Oak, Portsmouth, PO3 5PQ, United Kingdom}
\thanks{\ \ E-Mail: ronald.larham@baesystems.com}

\begin{document}


\begin{abstract}
In response to a challenge in a recent paper to measure the propagation speed of the wave of collapse of an 
array of dominoes (the Domino Effect), a novel method of measuring the 
speed of such waves has been developed using sound recordings of the collapse 
and DEMON (Detection of Modulation on Noise) analysis to extract the frequency of domino 
impacts and hence the speed of propagation of the domino wave. This paper presents this
method and a discussion of the other published measurements and models and some
comments on the precess of mathematical modelling.

\end{abstract}

\maketitle
\pagestyle{myheadings}
\markboth{R. Larham}{DEMONic Dominoes}


\section{Introduction} \label{Introduction}
\noindent Recent interest in the speed of propagation of the domino effect seems
to originate with a question by Daykin \cite{Daykin} in the problems section of
SCIAM review, and the initial response from McLachlan and Beaupre 
\cite{McLach} where they presented a dimensional analysis of the wave
speed and some experimental results.
Subsequently a number of authors have presented mathematical models of the
propagation of domino waves of varying levels of detail and complexity
(a partial list includes \cite{Efth} \cite{Stronge} \cite{StrongeShu}
\cite{StrongeBook} \cite{vanLee} \cite{Banks}).
 Also there have been additional measurements reported \cite{Stronge}
\cite{StrongeShu} (which may also be found in \cite{StrongeBook})
\\
\\
\noindent In a recent paper on the modelling of the propagation speed of domino waves 
\cite{Efth} a challenge was thrown down to actually measure the speed.
This seemed an interesting problem, and my initial thoughts were of videoing
the collapse of a domino array using a digital camera(a prime consideration
was that the experiment should have near zero impact on my household
finances so where possible use should be made of equipment that I already
owned or cost very little). After a start had been made on collecting
materials for the experiment and conducting some preliminary trials with
the dominoes, it occurred to me that the noise of the domino wave should
encode the frequency of dominoes impacting one another, and hence the speed
of the wave. As I has an old laptop computer with a sound recorder built in
and a spare computer microphone, recording the sound would entail zero
equipment cost, and would be less fiddly than videoing (extracting and
analysing the frames of a video is very time consuming, I know because
I have used the technique before when looking at the kinematics of bouncing).
\\
\\
All the results reported here used the same set/s of dominoes, their
dimensions are given in Section \ref{Results} .
\\
\\
What this paper does is demonstrate the application of some interesting
techniques of signal processing, some ideas from mathematical modelling
and in particular the need for model validation (that is the comparison
of model prediction with real data on the phenomena modelled to demonstrate
that at least for such cases the model is in acceptable agreement with the
model)


\section{Dimensional Theory} \label{DimensionalTheory}
\noindent McLachlan et al. \cite{McLach} conclude from dimensional analysis that the limiting
wave speed $V$  for thin dominoes satisfies:
\begin{eqnarray*}
 \frac{v}{\sqrt{gH}} = G(L/H) 
\end{eqnarray*}
for some function $G$. Which for thin dominoes is the same as:
\begin{eqnarray*}
 \frac{v}{\sqrt{gH}} = G_1(d/H) 
\end{eqnarray*}

\begin{figure}[htb]
\begin{center}
\includegraphics [width=4 in] {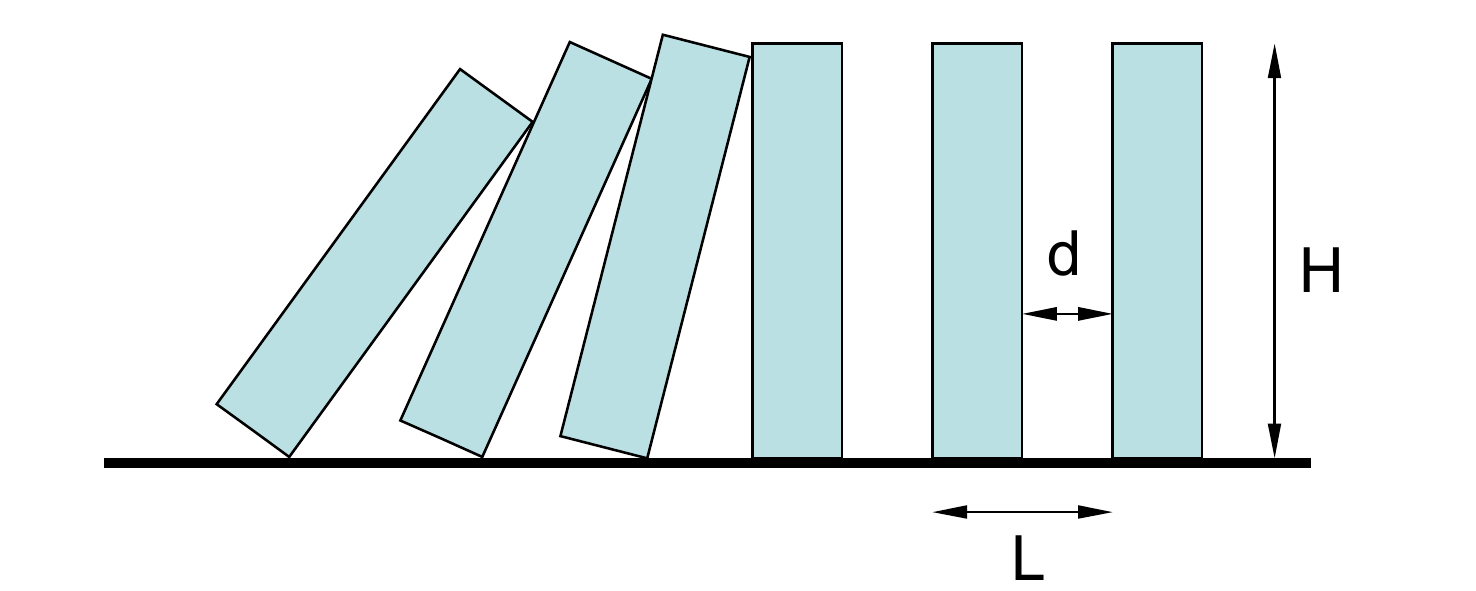}
\end{center}
\caption{Geometry of Domino Array} 
\label{geom} 
\end{figure}

\noindent Where $H$ is the height of the dominoes, $d$ the gap between adjacent
dominoes, and $L$ the distance between equivalent points on neighbouring
dominoes (that is the pitch of the domino array) (see figure \ref{geom} for
the significance of the variables). McLachlan et al. do not give their analysis that
leads to these results, but it is easy  enough to reconstruct.
\\
\\
Notes: There are additional dimensionless parameters hidden in functions $G$ and $G_1$
as the normalised speed also depend on the dimensionless constants
characteristic of the materials involved, in this case these include the coefficient of
friction between dominoes, and the coefficient of restitution for inter
domino impacts. The coefficient of friction between the surface and the dominoes
is of lesser relevance as in domino experiments it is usual to arrange things so that
there is no slipping between the dominoes and the surface. The models in
Stonge \cite{Stronge}, Strong and Shu \cite{StrongeShu} and Van Leeuwen
\cite{vanLee} represent the effects of these, but as the experimental results
show the material properties of the dominoes for the materials tested have
a minor influence on the wave speed. 

\section{Data Generation and Collection} \label{DataGeneration}
\begin{figure}[htb]
\begin{center}
\includegraphics[width=4.5 in]{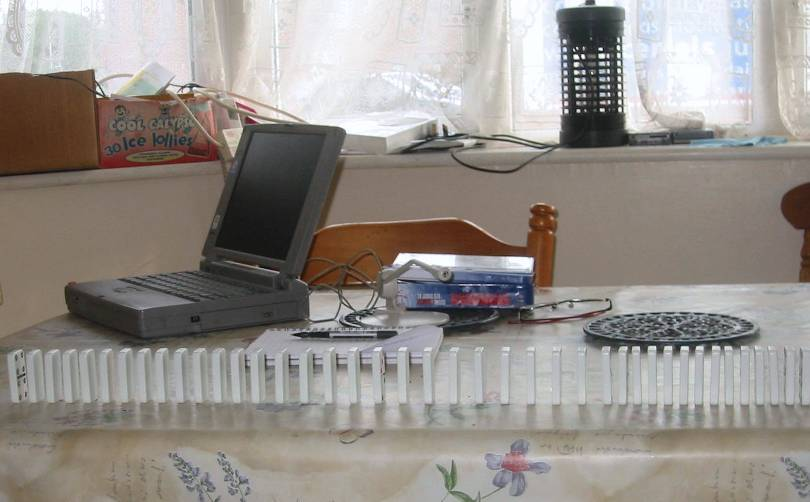}
\end{center}
\caption{Photo of Experimental Set Up for an Early Pilot Run} 
\label{pic} 
\end{figure}

\noindent Initially I had toyed with the idea of videoing domino waves, then
extracting the speed from an analysis of the video's frames. I 
abandoned this approach when I realised that audio recording would be
more convenient. The way that I decided to measure the domino wave speed
was to use the sound recorder and microphone on an old laptop to
record the sound of a domino array collapse. (This is far less demanding
in terms of cost of equipment than the high speed photography reported
in \cite{Stronge}, \cite{StrongeShu} and \cite{StrongeBook}). Then to analyse the recording
to extract the frequency of dominoes hitting the their adjacent domino (which is a simpler process than manually analysing frames of a video).
\\
\\
The experimental set up is shown in figure \ref{pic} (in any future experiments the
computer will be moved away from the rest of the set up as in retrospect it seems that
the computer fan was probably the limiting noise source for the experiments).
\\
\\

\noindent The signal of interest is encoded in the envelope of the recording so analysis
techniques analogous to the processing in a crystal AM radio
receiver, or a simple form of DEMON (Detection of Envelope Modulation
On Noise) analysis  similar to that used in passive Sonar processing is required
(unclassified references for DEMON, other than publicity releases for equipment that uses
it, are difficult to find but Kummert \cite{Kummert} includes a description).
The initial sections of each recording were progressively discarded to
identify and eliminate any start up transients. For most of the recordings
the transients were at most slight and easily eliminated, but four must be
regarded with caution (the two with the closest and the two with the widest
relatively spacing of the  dominoes) as the results for these were inconsistent
(they could be repeated more carefully).

\begin{figure}[htb]
\begin{center}
\includegraphics[width=5 in] {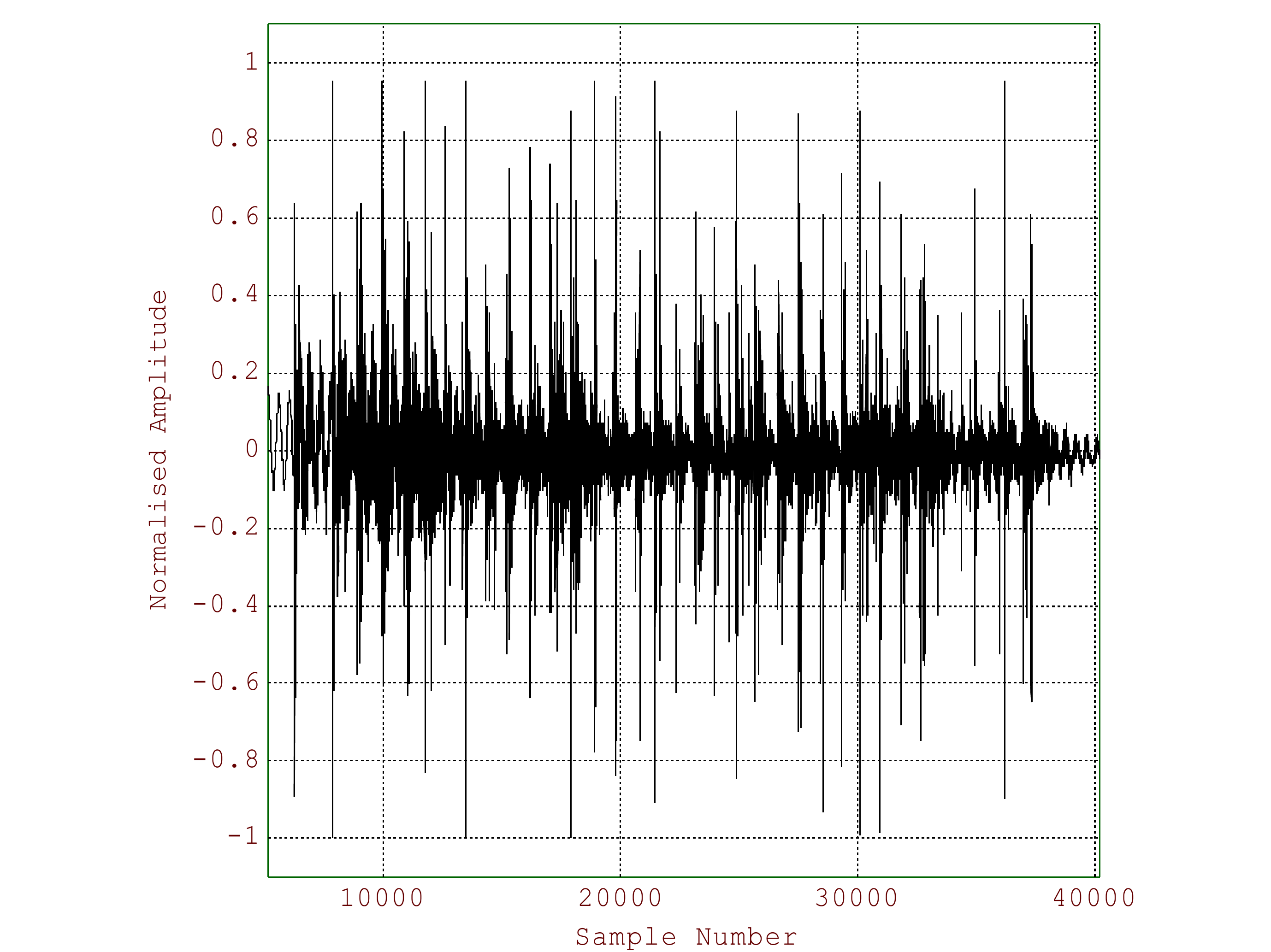}
\end{center}
\caption{Plot of the Sound Recording of a Domino Wave} 
\label{wave} 
\end{figure} 


\section{Processing of Acoustic Data} \label{Processing}
\begin{figure}[htb]
\begin{center}
\includegraphics[width=5 in] {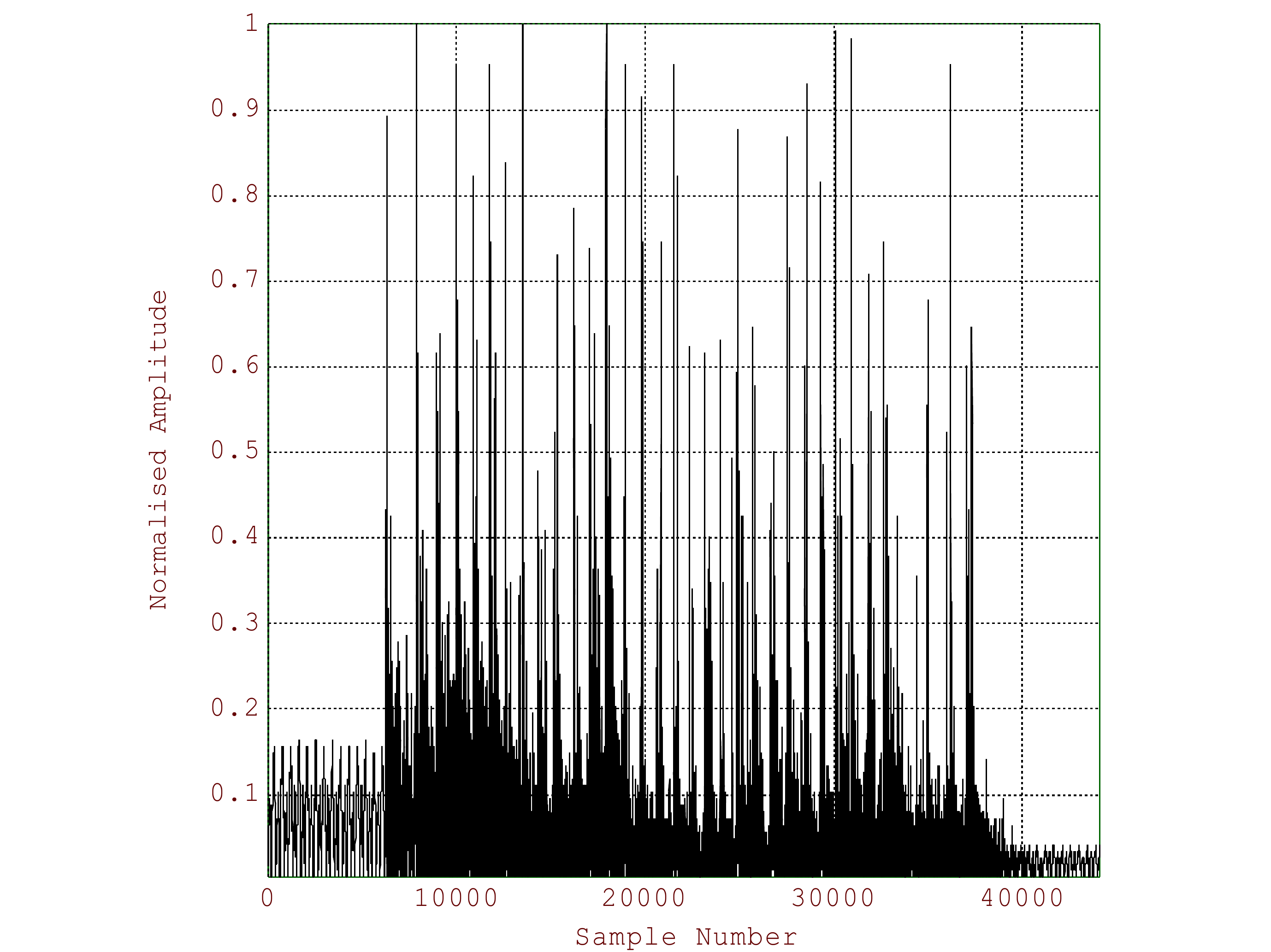}
\end{center}
\caption{Plot of Rectified Recording of Domino Wave} 
\label{rectfied} 
\end{figure}

\begin{figure}[htb]
\begin{center}
\includegraphics[width=5 in] {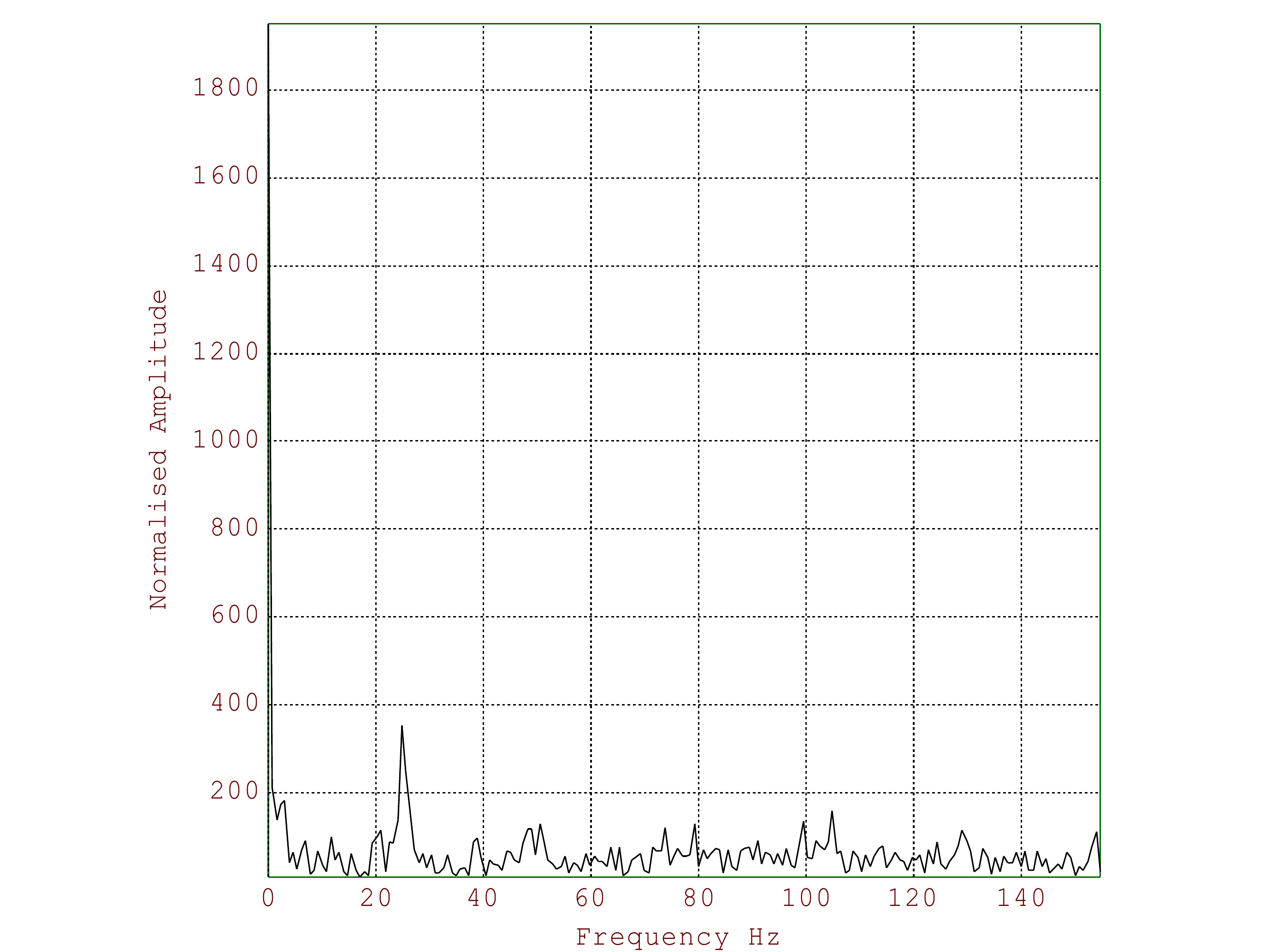}
\end{center}
\caption{DEMON Amplitude Spectrum of Signal} 
\label{DEMON1} 
\end{figure}

\begin{figure}[htb]
\begin{center}
\includegraphics[width=5 in] {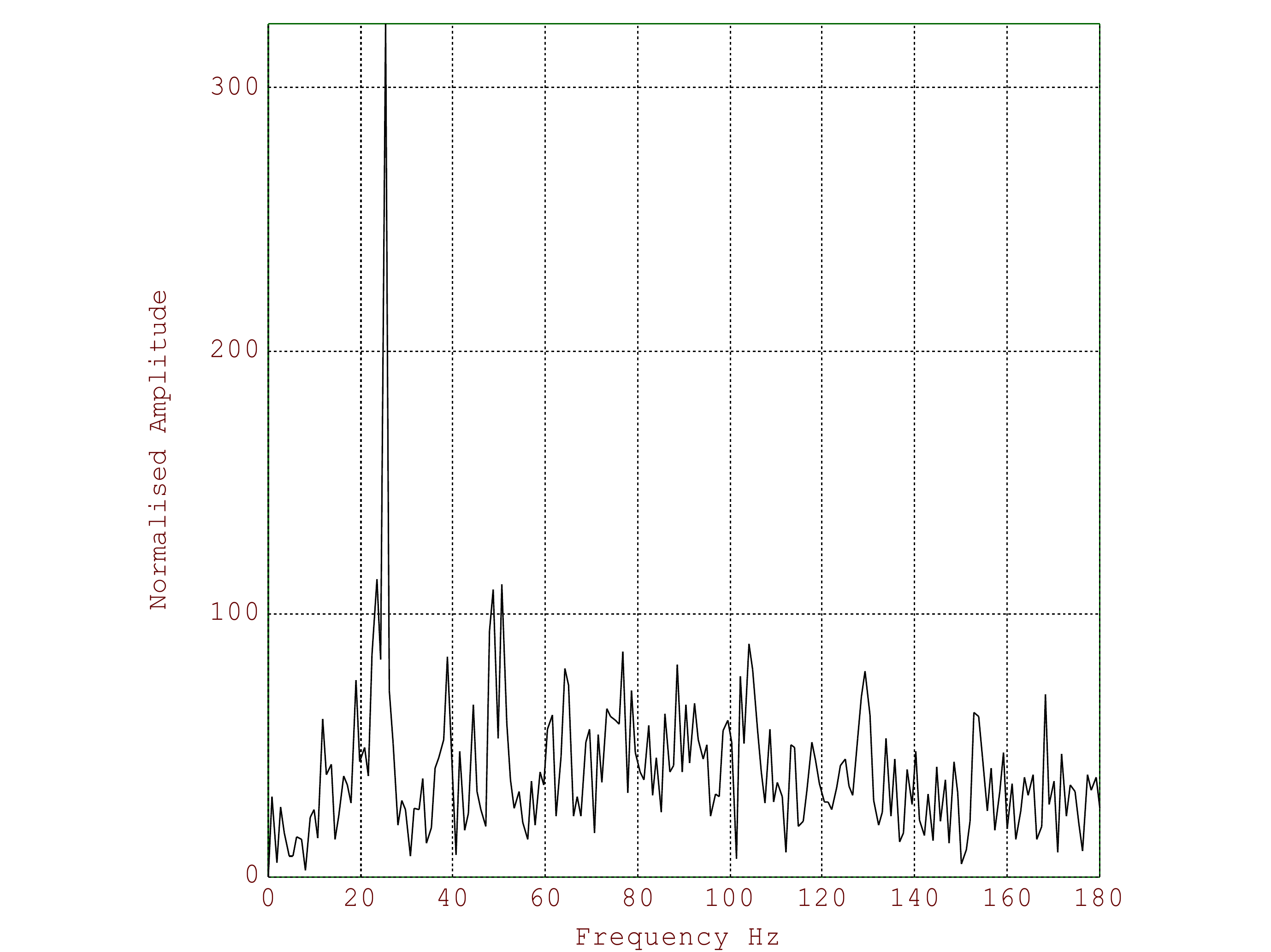}
\end{center}
\caption{Processed DEMON Amplitude Spectrum of Signal} 
\label{DEMON2} 
\end{figure}

\noindent The Windows sound recorder produces a .wav file as its output which contains 
the recorded data. This (in our case) was sampled at $\sim 22$ kHz (about
22000 samples per second) with 1 byte (8 bits) per sample, which in principle
gives $2^8\ (256)$ different levels. For analysis the data is shifted to
have zero mean and normalised to the range $(-1,1)$.
\\
\\
There are several artefacts in the recordings due to the way the sound recorder
operates, and the lack of controls on the version used.
The most conspicuous artefact is the result of the recorder's Automatic Gain
Control (AGC) which leads to the general decay
amplitude visible in figure \ref{wave} (The plots shown in figures \ref{wave}
- \ref{DEMON2} are for a domino array with $d/H=0.62$). Also just visible in
figure \ref{wave} is the zero offset in the short segment of data visible before
the sound of the dominoes starts to dominate. For the analyses that are applied
to the data these artefacts are of little to no importance, effectively introducing 
additional ``noise" which we will see is not a real problem. 
\\
\\
Looking at figure \ref{wave} or the plot of the rectified data shown in figure \ref{rectfied} we see a series of spikes that
look as though they are near periodic, these are predominantly the clicks of
the dominoes hitting one another. It is the average frequency of occurrence
of these clicks together with the nominal domino spacing that allows us
to deduce the speed of the domino wave.
\\
\\
In order to extract the ``average" frequency of the spikes we perform a frequency
analysis of the rectified waveform shown figure \ref{rectfied}. We use the rectified
data for this because the spectrum of the unrectified data shows no obvious features
at the spike frequency, the dominant low frequency feature is hum at around 50 Hz. The
spike frequency if present will appear as a modulating frequency on tones (of phase
random from spike to spike) or on noise. If we look at the full spectrum our suspicion
that there will be no features that are easily identifiable as such are confirmed.
Figure \ref{DEMON1} shows the low frequency part of the spectrum of the rectified signal.
Here we see a large spike at zero frequency due to the positivity of the signal, the next 
peak at $\sim 25$ Hz is the frequency we seek, there are also faint signs of harmonics
of this frequency (these are more obvious in equivalent plots for some of the other
domino spacings). We also see that the hum (which should now appear at $\sim 100$ Hz)
is small compared to the feature of interest. That the feature identified in 
figure \ref{DEMON1} corresponds to the spike spacing in figure \ref{rectfied} can be
shown by measuring the spacing of the spikes in figure \ref{rectfied}.
\\
\\
The use of the FFT algorithm to perform the required frequency analysis is
discussed in Appendix A.
\\
\\
The above explains the main ideas of our analysis, but to make the feature of
interest clearer we filter to a band that includes the majority
of the energy in the spikes and also filter out the low frequency components below
$\sim 10$ Hz after rectification. This gives us the much clearer signature shown
in figure \ref{DEMON2}. It is these plots of the processed data that I use to take 
measurements from. The data in this paper was extracted from such plots essentially
by measuring semi-manually from such plots. This could be automated, and the centroid
of the peaks computed rather than manually measuring the position of the tip of the peak,
but I have not done that for this paper. 

\section{Results} \label{Results}

\noindent The experiments were all conducted with dominoes of dimensions $\approx 0.0516 \times 0.0255 \times 
0.0079$ meters. The results shown in table \ref{table1} are for dominoes with a vertical 
orientation (standing on their smallest faces), and those in table \ref{table2} are for
dominoes with a horizontal orientation (standing on their second smallest face).
\\
\\

\begin{table}[htb]
\begin{center}
\caption {Experimental Results With Dominoes Vertical (italic script indicates less reliable data)}
\begin{tabular}{|c|c c c c c c c c c|}
\hline
$d/H$ & 0.04 & 0.14 & 0.23 & 0.33 & 0.43 & 0.53 & 0.62 & 0.72 & 0.82\\
\hline
$V/\sqrt{gH}$ & \sl{1.07} & \sl{1.33} & 1.53 & 1.51 & 1.47 & 1.50 & 1.40 & 1.33 & \sl{1.23}\\
\hline
\end{tabular}
\label{table1}
\end{center}
\end{table}

\begin{table}[htb]
\begin{center}
\caption {Experimental Results With Dominoes Horizontal (italic script indicates less reliable data)}
\begin{tabular}{|c|c c c c|}
\hline
$d/H$ & 0.28 & 0.47 & 0.67 & 0.87\\
\hline
$V/\sqrt{gH}$ & 1.15 & 1.19 & 1.15 & \sl{0.68}\\
\hline
\end{tabular}
\label{table2}
\end{center}
\end{table}

\begin{figure}[htb]
\begin{center}
\includegraphics[width=5 in] {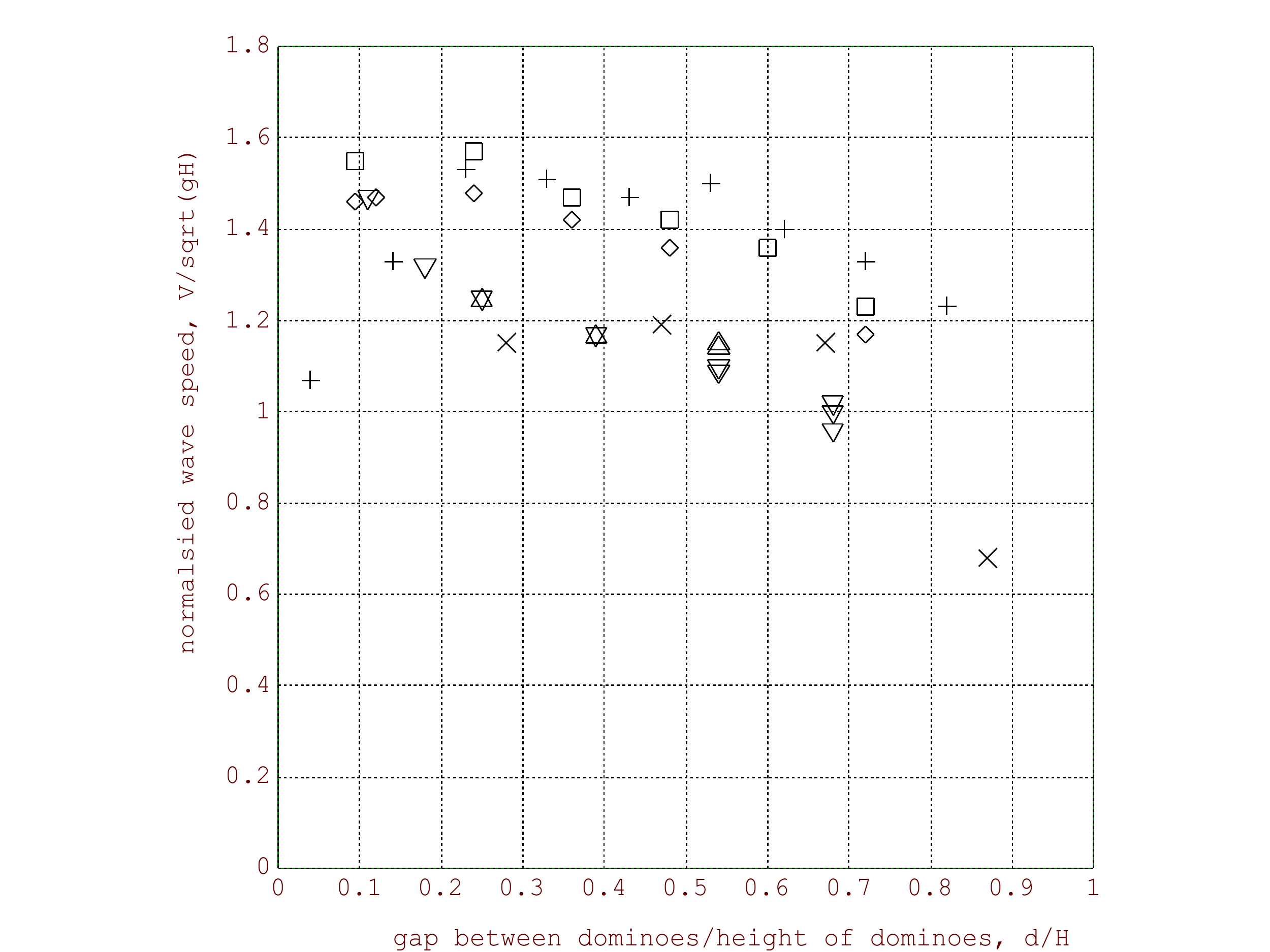}
\end{center}
\caption{Domino Wave Speed Data;
$+$ measurements by the current author with the dominoes vertical, $\times$
with them horizontal, $\square$ and $\lozenge$ measurements from Strong and Shu 
\cite{StrongeShu}, $\triangledown$ (normal dominoes) and $\vartriangle$ (double
height dominoes) measurements from McLachlan et al.
\cite{McLach} } 
\label{graph2} 
\end{figure}

\begin{figure}[htb]
\begin{center}
\includegraphics[width=5 in] {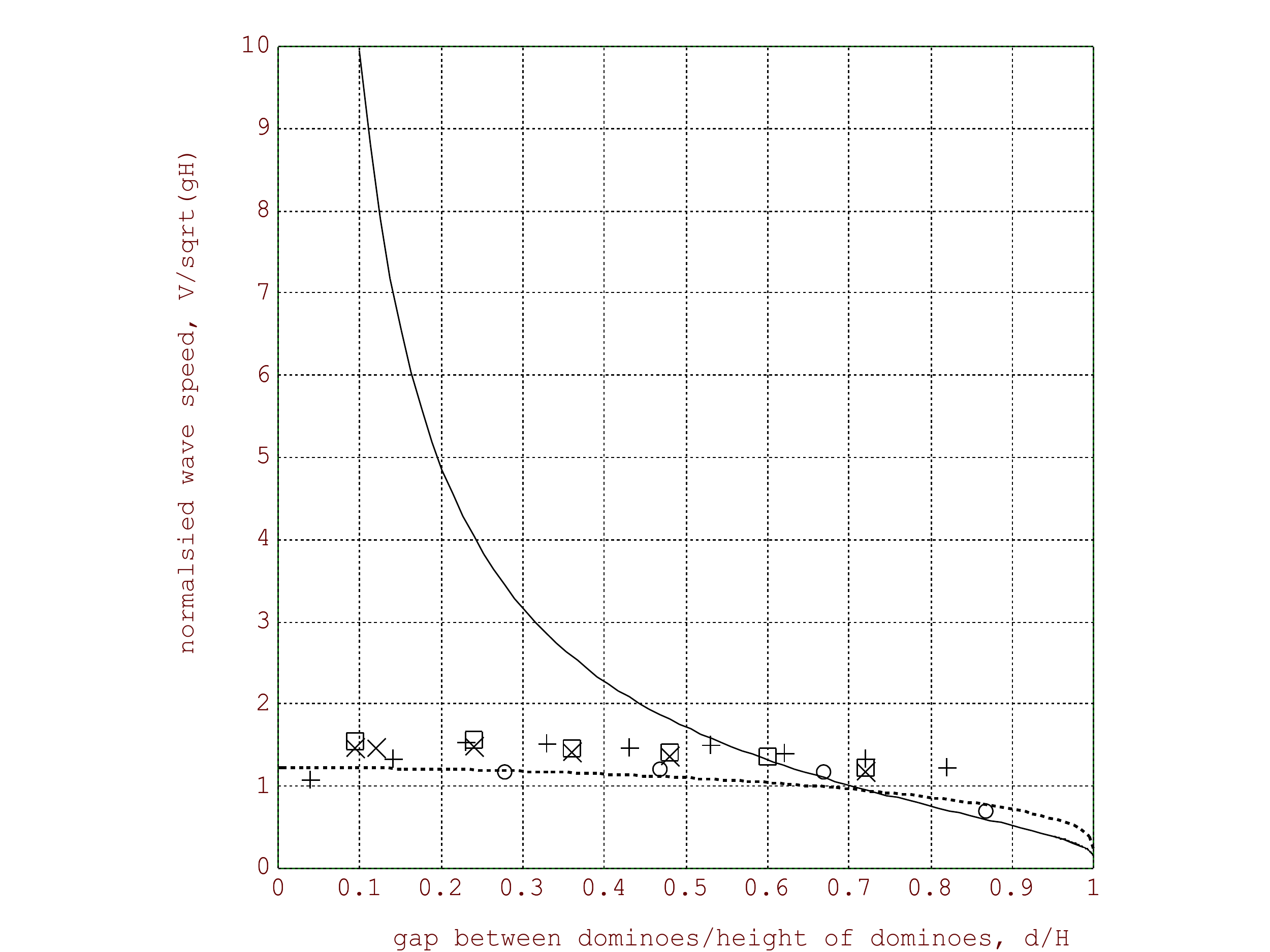}
\end{center}
\caption{Domino Wave Speed Plot; the solid line is the model prediction from
Efthimiou and Johnson \cite{Efth}, dashed line from Banks \cite{Banks},
$+$ measurements by the current author with the dominoes vertical, $\circ$
with them horizontal, $\times$ and $\Box$ measurements from Strong and Shu
\cite{StrongeShu} } 
\label{graph1} 
\end{figure}

\noindent \textbf{Notes}
The last entry in table \ref{table2} has a spacing greater than the maximum
for which one would expect the domino wave to propagate. At a value of $d/H > \sqrt{3}/2$
a domino strikes its neighbour below its' mid point, and under these conditions
it may well not topple in the expected manner, this is van Leeuwen's practical upper limit
for the wave to propagate. So it is no surprise that
the data for this point is unreliable and this was the largest spacing at which 
I could get the wave to propagate. Presumably it did propagate in this case
as a result of the irregularities in the domino geometry and spacing, or some
other unidentified reason.
\\
\\
All of the papers that report domino wave speed measurements
report speeds $\sim 0.9$ to $1.7 \times \sqrt{gH}$. These are in
broad agreement with my own measurements, my and 
other published measurements are shown in figure \ref{graph2}.
\\
\\
As the results shown in table \ref{table2} are systematically 
lower that those in table \ref{table1} so we may suspect that
one or more assumptions underlying the dimensional analysis
are invalid.
\\
\\
Given the usual shapes of dominoes I would hope that the thin domino
approximation would be not unreasonable down to values of $d/H \sim 0.2$.
\\
\\
As can also be seen in figure \ref{graph1} the measured data are comparable to
the predictions of \cite{Efth} over a rather limited range of
$d/H$. This is in contradistinction to the models the predictions of Bank's
\cite{Banks} which in general give rather better agreement with experiment.
The models which represent the effects of multiple dominoes being involved in the
collapse wave being rather better than Bank's model. Even so the
reasonable agreement between the experimental data and the model
predictions from Banks \cite{Banks} is worth noting as it indicates that
the single neighbour domino interaction assumption is not entirely misleading.
\\
\\
The spectral features corresponding to the wave speeds are often split into two or
three closely spaced features (typically a few Hertz apart). This may be due to 
irregularities of the surface used for the experiments, or to some irregularity
in the dominoes. When checked with a spirit level the table surface appears to be
flat, but close examination of the dominoes seems to indicate that opposite
short edges are not parallel. The irregularity in the dominoes appears to be
substantially the same for all the dominoes, and so may be responsible for the
splitting of the spectral features.
\\
\\
To gain some idea of the errors associated with the better data points the
domino wave speed was measured multiple times for one value of domino spacing
and the mean and standard deviation or the wave speed computed. This give
the result that for $d/H=0.62$ we have a mean non-dimensional wave speed
$V/\sqrt{gH}$ of $1.37$ with   standard deviation estimated from the
sample of $\sim 0.07$. 

\section{Discussion} \label{Discussion}
\noindent The experimental data may be summarised as telling us that to a fair
(hand-waving) approximation for common dominoes the normalised wave
speed is a relatively weak function of the normalised inter domino interval for
practical intervals (or at most shows a slight downward trend with increasing
domino spacing). Also that the normalised  wave speed is of the $\sim 1$.
\\
\\

From figure \ref{graph2} we can see that all the reliable data points
measured in this study give normalised wave speeds in the range $\sim 1 - 1.6$
which is in reasonable agreement with other measurements.
\\
\\
It could be interesting to do some further work to improve the measurements
for closely spaced dominoes with $d/H \sim 0.05-0.2$ as the current data is
poor here but may with more careful work be capable of improvement. This would
be interesting even if only to see how far the technique can be pushed. It would
also be worthwhile to see if the quality of all the data can be improved by
being careful to arrange for all the dominoes to have the best orientation.

\section{Summary}\label{Summary}
\noindent From the comparison of the model of Efthimiou and Johnson \cite{Efth} and 
experiment we see that the area of agreement of experiment and model is rather limited.
Had the model been part of a project with some economic impact we would have been
at risk of being found to not have shown due diligence, which could result in
unfavourable consequences for us and/or our employers in the event of
a failure.
\\
\\
Validation of models is not a chore that we may do after the interesting
parts of a study are completed but an essential activity if our work
is not to be nugatory.
\\
\\
It is also worth while comparing the predictions in the literature with ones
current models predictions, the differences may be important and in need of
explanation

\newpage
\section*{Acknowledgements}
\noindent The author thanks his partner and children for putting up
with domino experiments on the kitchen table extending over
many evenings and weekends without expressing any more than slight
derision (or interest), the cats for not prematurely disturbing 
the domino arrays.


\appendix
\section{The Fast Fourier Transform and Frequency Analysis}
\noindent When doing a frequency analysis we want to look for significant
frequencies in the given signal. To do this we look at the frequency spectrum
of the signal which  is the square absolute value or just the absolute value
(or amplitude) of the Fourier Transform (FT) of the
signal. The FT breaks our signal $x(t)$ down into a linear combination of 
sinusoidal components, where the component at frequency $f$ is given
by:

\begin{displaymath}
X(f)=\mathfrak{F}x|_{f}=\kappa \int_{-\infty}^{\infty}x(t) e^{-i(2\pi f t)}\ dt
 \ \ \ ...(1)
\end{displaymath}

\noindent where $\kappa$ is a normalising factor the value of which I will not worry 
about as every area of application of the FT uses a different convention
for normalising factors. Also the negative sign in the exponential term
may in some versions of the FT be a positive sign, but none of this matters 
for what I am going to do, also at some point I will use library software and I 
don't want to have to worry about the conventions in use, if necessary 
I will normalise the spectrum to have the same energy as the signal (that is the
normalisation will make the integrals of their square magnitudes equal) . There is
an additional ambiguity in the definition of the FT and that is over the use
of angular frequency $\omega$ or plain frequency $f$, for now I will stick with $f$.
\newline
\newline
\noindent Now because we have a finite recording of the signals of interest
the range of integration may be reduced to a finite interval which contains
the recording:

\begin{displaymath}
X(f)=\mathfrak{F}x|_{f}=\kappa \int_{a}^{b}x(t) e^{-i(2\pi f t)}\ dt \ \ \ ...(2)
\end{displaymath}

\noindent which is now equivalent to the computation of the coefficients of
a Fourier Series and all the information in $X(f)$ is contained in
$X(n/(b-a)),\ n=0,\ \pm 1,\ \pm 2...$ (in fact since $x(t)$ is a real
signal $X(f)$ has complex conjugate symmetry and so everything about
$x(t)$ is encoded in  $X(n/(b-a)),\ n=0,\  +1,\ +2,\ ...$)
\newline
\newline
There are several problems with $(2)$ but the main one is that while the actual
signal of interest is a function of the continuous time variable
$t$ we only know its value at discrete sample points. To get around
this problem we can use a numerical integration scheme to compute
the Fourier coefficients. The scheme that I adopt is the simplest
so I approximate:

\begin{displaymath}
X(k/(b-a))\approx \kappa \sum_k x(t_k) \Delta t e^{-i(2 \pi n t_k/(b-a))}\ \ \ ...(3)
\end{displaymath}

\noindent where $t_k=a+k \Delta t,\ \ k=0,\ ...,
\ \lfloor (b-a)/\Delta t \rfloor$ which with a bit of jiggery pokery 
will allow the use of Fast Fourier Transform (FFT) algorithms
to do our computations (for simplicity we will generally work with $b-a$
being an integer multiple of $\Delta t$). This is a desirable result
because of the almost incredible efficiency of FFT algorithms. This approach
is known to work well if the signal has negligible energy at frequencies
above half the sampling rate (Nyquist frequency or rate) used,
which is usually the case as the recording hardware will generally 
filter the signal to the required band before sampling.
\\
\\
\section{Data Analysis Software}
\noindent The data analysis package used to to the processing described in this note was
Euler Math Toolbox (EuMathT) (or rather my version of an earlier incarnation simply called 
Euler) This required minor modifications to the function for reading wav files to correct
for a difference in the file format produced by the recorder and that expected by
the function, the current version of EuMathT may have fixed this problem.
In addition to the built in facilities of the data analysis package additional
code to do the specific analysis and plotting required is written in the packages
own BASIC like matrix language.
\\
\\
Alternatives to EuMathT exist and any of them would be an equally suitable tool for
this job. The obvious commercial alternative is Matlab, with other freeware or Open
Source packages being: SciLab, FreeMat, Octave (which to a greater of lesser extent
use syntax compatible with Matlab) and Yorick. Some of the alternatives may need tweaking
to get them to read wav files, but generally if the package does not have this facility
built in code to implement it can usually be downloaded from the Internet. A simple
search with your favourite Internet search engine will turn up links to the websites for
all of these packages.

\end{document}